# Effect of the Three-Site Hopping Term on the $t$-$J$ Model

B. Ammon[a,b], M. Troyer[a,b], Hirokazu Tsunetsugu[a,b]
[a]*Interdisziplinäres Projektzentrum für Supercomputing,
Eidgenössische Technische Hochschule, CH-8092 Zürich,*
[b]*Theoretische Physik, Eidgenössische Technische Hochschule,
CH-8093 Zürich, Switzerland*
(February 9, 1995)

We have used exact diagonalization and quantum Monte Carlo methods to study the one-dimensional $t$-$J$ model including the three-site hopping term derived from the strong coupling limit of the Hubbard model. The three-site term may be important to superconducting correlations since it allows direct hopping of local singlet electron pairs. The phase diagram is determined for several values of the strength of the three-site term and compared with that of the $t$-$J$ and Hubbard models. Phase separation, which exists in the $t$-$J$ model is suppressed. In the low electron density region the formation of local singlet electron pairs is enhanced, leading to stronger superconducting correlations even for values $J/t < 2$. A large spin gap region extends from low electron densities up to high densities. In the low hole density region the superconducting correlations are suppressed at $J/t > 2.8$ in spite of enhanced pair formation. This is because the three-site term, while enhancing the formation of electron pairs, leads to a repulsion between holes.
74.20.Mn, 71.27.+a, 75.10.Jm, 74.25.Dw

## I. INTRODUCTION

Since the discovery of high-$T_c$ superconductors,[1] the investigation of strongly correlated fermion systems is a central theme of theoretical solid state physics. It has been established that the relevant physics of these cuprate superconductors is contained in the two dimensional (2D) $CuO_2$ planes. Anderson suggested that a single-band Hubbard model would be the minimal model needed to describe the low energy properties.[2] The Hamiltonian can be written as

$$H_{\text{Hubbard}} = -\sum_{\langle i,j\rangle,\sigma} t c_{i\sigma}^\dagger c_{j\sigma} + U\sum_i n_{i\uparrow}n_{i\downarrow}, \qquad (1)$$

where $\langle i,j\rangle$ denotes a pair of nearest neighbors, $c_{i\sigma}^\dagger$ and $c_{i\sigma}$ are the usual electron creation and destruction operator at the site $i$ with spin $\sigma$ and $n_{i\sigma} = c_{i\sigma}^\dagger c_{i\sigma}$. The on-site Coulomb term $U$ describes strongly correlated electrons in the limit $U/t \to \infty$ as well as weakly coupled electrons for $U/t \ll 1$.

Strong correlations make the limit $U/t \to \infty$ especially interesting but also difficult to investigate. The system avoids doubly occupied sites in this limit, since they cost a large energy $U$. In addition to the local constraint of excluding double occupancy, interactions take place through virtual double occupancy. By a canonical transformation we get an effective Hamiltonian[3,4] which reads to second order in $t/U$:

$$\begin{aligned}H_{t\text{-}J\text{-}t_3} &= -\sum_{\langle i,j\rangle\sigma} t\left(\hat{c}_{i\sigma}^\dagger \hat{c}_{j\sigma} + \text{h.c.}\right) \\ &+ J\sum_{\langle i,j\rangle}\left(\vec{S}_i\cdot\vec{S}_j - \tfrac{1}{4}n_i n_j\right) \\ &- t_3 \sum_{\langle i,j,k\rangle\sigma}\left(\hat{c}_{k\sigma}^\dagger n_{j-\sigma}\hat{c}_{i\sigma} - \hat{c}_{k\sigma}^\dagger \hat{c}_{j-\sigma}^\dagger \hat{c}_{j\sigma}\hat{c}_{i-\sigma} + \text{h.c.}\right).\end{aligned} \qquad (2)$$

This model which we call $t$-$J$-$t_3$ model is the subject of this paper. We set $J = 4t^2/U$ and $t_3 = \alpha J/4$. The value $\alpha = 1$ is obtained by the canonical transformation, but we will also consider smaller values of $\alpha$ to provide a continuous switch on of the pair hopping terms. The brackets $\langle i,j,k\rangle$ denote a three site term, with $i \neq k$ being neighbors of $j$. The pseudo-electron creation operator $\hat{c}_{i\sigma}^\dagger = c_{i\sigma}^\dagger(1 - n_{i-\sigma})$ allows only single occupancy of each site, $n_i = n_{i\uparrow} + n_{i\downarrow}$ and $\vec{S}_i$ denotes the spin operator. Usually the pair hopping terms are neglected,[4,5] and we obtain the familiar $t$-$J$ model for $\alpha = 0$. We note that it has been shown by Zhang and Rice that the $t$-$J$ model can also be derived from a three-band model,[6] describing the $CuO_2$ planes.

We can write the interaction term in $H_{t\text{-}J\text{-}t_3}$ alternatively as a number operator and a hopping term for singlet pairs of electrons:

$$H_{\text{int}} = -J\sum_i P_i^\dagger P_i - \tfrac{1}{2}\alpha J\sum_i (P_i^\dagger P_{i+1} + \text{h.c.}), \qquad (3)$$

where $P_i^\dagger = (1/\sqrt{2})(c_{i\uparrow}^\dagger c_{i+1\downarrow}^\dagger - c_{i\downarrow}^\dagger c_{i+1\uparrow}^\dagger)$ is the creation operator of a nearest neighbor singlet pair. This allows electron pairs to hop without the cost of breaking a bond $J$. We expect therefore at least in the low electron density region a strong stabilization of local electron singlet pairs. If this stabilization is strong enough, we will get a short range resonating valence bond (RVB) state, forming a gas of local singlet pairs. The lowest spin excitation of this short range RVB state will then be the excitation of a local singlet pair to a triplet state, and hence have a finite spin gap. An important difference to the usual $t$-$J$ model ($\alpha = 0$) is that the local singlet electron pairs of this short range RVB phase can gain kinetic energy of the order of $\alpha J$ due to the pair hopping term and we may expect an expansion of this spin gap phase to higher electron densities. Since a finite spin gap is characteristic of this state, we will also investigate the region of a finite spin gap the $t$-$J$-$t_3$ model in this paper.

Also related to the kinetic energy is the occurrence of phase separation. The actual structure of the electron rich and poor phases is determined by a compromise of a gain of kinetic energy and a cost of exchange energy, and therefore depends on the ratio $J/t$ and the lattice structure. In the one-dimensional (1D) $t$-$J$ model, the phase separation starts at $J/t \sim 2.8$ (for small electron density) and 3.6 (near half filling).[7,8] On the other hand, the critical $J/t$ value for phase separation is smaller near half filling in two dimensions.[9] The phase separation in the $t$-$J$-$t_3$ model would be quite different from the $t$-$J$ model. This is because even for large $J/t$ the system now can gain a large kinetic energy by pair hopping and consequently the phase separation is expected to be strongly suppressed.

To study the effects of this new pair hopping term in detail, we have limited our calculations to the 1D case for simplicity.

## II. NUMERICAL METHODS

Two complementary numerical algorithms have been used to investigate the properties of these systems, exact diagonalization of small clusters by the Lanczos algorithm[10,11] and the Quantum Monte Carlo (QMC) world line algorithm.[12]

A careful choice of the boundary conditions is important for exact diagonalization. One reasonable choice is the boundary condition where all one electron orbitals in the noninteracting case are either fully occupied or empty (closed shell boundary conditions, CSBC).[7,13,14] For 1D single-band models, this corresponds to periodic boundary conditions (PBC) for systems with $N = 4m+2$ electrons and antiperiodic boundary conditions (ABPC) for $N = 4m$, with $m$ being an integer. With this boundary condition the ground state is a spin singlet. Alternatively choosing the opposite boundary conditions, APBC for $N = 4m + 2$ and PBC for $N = 4m$ will be called open shell boundary conditions (OSBC). If not otherwise stated, CSBC is used for all calculations.

For the QMC calculations, a parallel version[15] of the world line algorithm with a four site cluster decomposition[16] has been used. The advantage of QMC methods is that they allow simulations of considerably larger systems. We have calculated systems of up to 64 sites at an inverse temperature of $\beta t = 32$, since the negative sign problem did not appear for these models. At such low temperatures, the measured quantities do not change qualitatively compared to zero temperature. The simulations have been restricted to the subspace of zero winding number.[8] To control the systematic error of order $O(\Delta \tau^2)$ due to the finite Trotter time step $\Delta \tau$, measurements have been performed at $\Delta \tau t = 0.25$ and $0.1$ and extrapolated to $\Delta \tau t = 0$.

## III. PHASE SEPARATION

The ground state of strongly correlated electron systems is not always homogeneous, but sometimes is phase separated, consisting of a mixture of two phases. The phase separation line divides the region of stable homogeneous phases in the parameter space from the mixed phases region. Whereas the repulsive Hubbard model is homogeneous for the whole parameter range,[17–19] the $t$-$J$ model is phase separated for large $J/t$.[7,8]

The onset of phase separation is estimated as a divergence of the compressibility. We used a finite size approximation

$$\kappa = \frac{L}{N^2} \frac{4}{E_0(L; N+2) + E_0(L; N-2) - 2E_0(L; N)}, \quad (4)$$

with $E_0(L; N)$ being the ground state energy of a system with $N$ electrons in $L$ sites.

For the low electron density region, the phase separation boundary can be estimated by an argument of Emery, Kivelson, and Lin.[20] We start with a Heisenberg chain of length $N \gg 1$ in a phase separated state. Now we assume that two electrons evaporate from the chain to form a singlet bound state at a large distance from the chain, and compare the energy of this state with the initial state energy. Solving the two electron problem, it is found that a singlet bound state appears if

$$J > J_{\text{pair}} = \frac{2t}{1+\alpha}, \quad (5)$$

and the energy of this state is

$$E_{\text{pair}} = -(1+\alpha)J - \frac{4t^2}{(1+\alpha)J}. \quad (6)$$

Evaporating the two electrons costs the energy of two bonds, $2J \ln 2$. Phase separation starts when $E_{\text{pair}} > -2J \ln 2$ or at $J > J_{\text{cr}}$ with

$$J_{\text{cr}} = \frac{2t}{\sqrt{(2 \ln 2 - (1+\alpha))(1+\alpha)}} \quad (7)$$

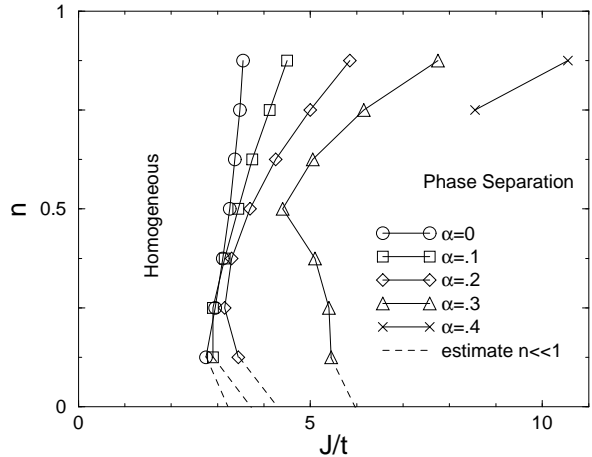

FIG. 1. Boundary line of phase separation in the $t$-$J$-$t_3$ model for different values of $\alpha$, calculated by exact diagonalization of a lattice of length $L = 16$ and CSBC. The phase separation line is shifted towards larger $J/t$ with increasing the pair hopping term and disappears completely.

TABLE I. Three site correlation of two particles and one hole $p_{sh} = \langle n_i h_{i+1} n_{i+2} \rangle/(1-n)$ for a density $n = 7/8$ in the $t$-$J$-$t_3$ model, calculated by exact diagonalization on a lattice of length $L = 16$.

| | $J/t$ | | | | |
|---|---|---|---|---|---|
| $\alpha$ | 1 | 2 | 3 | 5 | 10 |
| 0.0 | 0.984 | 0.970 | 0.926 | 0.389 | 0.064 |
| 0.4 | 0.982 | 0.971 | 0.958 | 0.927 | 0.844 |
| 1.0 | 0.980 | 0.970 | 0.962 | 0.947 | 0.923 |

in the low electron density limit. This argument assumes $J_{cr} > J_{pair}$, which is always fulfilled. The estimate for $J_{cr}$ would be exact if there are no bound states with more than two electrons. Following this argument, phase separation disappears at low electron densities for $\alpha \geq \alpha_{cr} = 2\ln 2 - 1 \approx 0.386$. In Fig. 1 we have plotted the phase separation boundaries in the phase diagram defined in the plane of the electron density $n = N/L$ and the coupling $J/t$ for different values of $\alpha$. The figure shows that at $\alpha = 0.4$ the phase separation region survives only near $n = 1$ in agreement with the above argument. We show only two points for the phase separation line at $\alpha = 0.4$, since here the phase separation line diverges very rapidly to infinity for $n < 3/4$.

We can see in Fig. 1 that also in the higher electron density region the onset of phase separation is pushed away towards larger values of $J/t$ with increasing $\alpha$. At $\alpha = 1$ phase separation completely disappears for all densities $n$. We can see in Fig. 4(b) that the compressibility $\kappa$ is strongly reduced for $\alpha = 1$ in contrast to the divergence in $\kappa$ for $\alpha = 0$ and 0.4. This point will be discussed in more detail in Sec. V. Very small values of the compressibility $\kappa$ in the low hole density region may indicate a repulsion among holes. We can see this more explicitly by by considering the probability of finding a single hole between two electrons, $p_{sh} = \langle n_i h_{i+1} n_{i+2} \rangle/(1-n)$. A value of $p_{sh}$ close to one indicates that practically all holes are isolated, while $p_{sh} \to 0$ suggests that the holes attract each other. The results in the low hole density region shown in Table I indeed show that $p_{sh}$ becomes close to one for $\alpha = 1$, and demonstrate a repulsion among holes. This effective repulsion is produced by the large gain of energy of the order of $\alpha J$ due to pair hopping, rather than the strongly renormalized small direct hopping. In the low hole density region, pair hopping enhances propagation of single holes, because it can gain more kinetic energy than propagation of bound hole pairs, and hence produces an effective repulsion among holes. This strong repulsion is the reason of suppression of phase separation.

Attention has to be paid on the shape of the phase separation line. The Maxwell construction shows that the phase separated region consists of a mixture of the two homogeneous phases that are on the crossing points of the phase separation line and the line of constant coupling $J/t$. In $t$-$J$ model, the phase separation line crosses the line of constant $J/t$ only once. In this case, the electron poor phase always is a vacuum. In the $t$-$J$-$t_3$ model, the phase separation line crosses the line of constant $J/t$ twice at certain values of $\alpha$ and $J/t$. Then the electron poor part is a dilute gas of bound pairs and the electron rich phase is a Heisenberg chain with a finite density of holes inside.

## IV. SPIN GAP

The spin gap is the excitation energy from the singlet ground state to the lowest triplet state $\Delta_s = \lim_{L\to\infty} \Delta_s(L; N = Ln)$, where $\Delta_s(L; N) = E_0(L; N; S^z = 1) - E_0(L; N; S^z = 0)$. The determination of the spin-gap region is an important issue in the characterization of a model with spin degrees of freedom. As it was stated already in the introduction, the presence of a spin gap can indicate a short-range RVB mechanism.

A finite spin gap is present in several extended $t$-$J$ models. One example is the $t$-$J$-$J'$ model, where a next nearest neighbor Heisenberg interaction $J'$ is added. It has a finite spin gap in the low hole doping region.[21] Electron pairing is also favored by adding a repulsive next nearest neighbor interaction $V$ and this effect introduces a finite spin gap in the $t$-$J$-$V$ model at quarter filling.[14] One common characteristic of these models is the restriction of the spin gap to the vicinity of the phase separation line. A small spin gap region exists also for the $t$-$J$ model in the low electron density region at $2.0 < J/t < 2.95$.[7]

The spin gap region for the $t$-$J$-$t_3$ model is determined for densities of $n = 1/2$, $2/3$ and $4/5$ by exact diagonalization of chains with different sizes. One needs to use CSBC and OSBC as explained in Sec. II. We have used

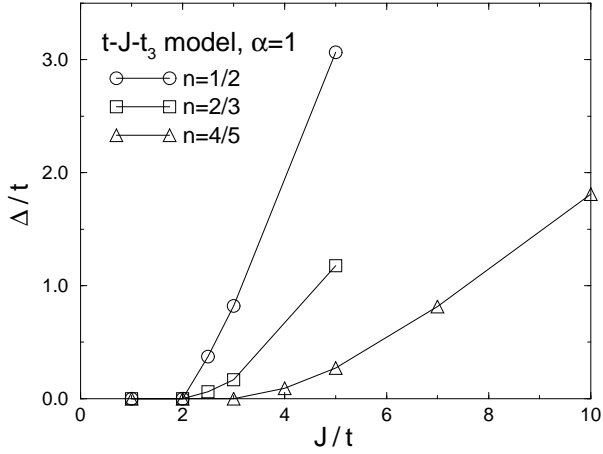

FIG. 2. Extrapolated value of the spin gap for the $t$-$J$-$t_3$ model with $\alpha = 1$ at several densities as a function of the coupling $J/t$.

the same fitting function as Refs. 14 and 21 for extrapolation to $L \to \infty$:

$$\Delta_s^{\rm BC}(L) = \Delta_s + \frac{a_1^{\rm BC}}{L} + \frac{a_2^{\rm BC}}{L^2}, \qquad (8)$$

where BC denotes the boundary condition.

In the low electron density region, the spin gap region is estimated to start where a singlet bound state of electron pairs is formed, that is at $J = J_{\rm pair}$ (Eq. (5)). With increasing $\alpha$, $J_{\rm pair}$ decreases, and the onset of the spin gap region shifts therefore towards smaller couplings $J/t$ for $n \ll 1$. For $\alpha = 0.4$ and 1 we have found numerically that the spin gap region extends up to densities of $n \gtrsim 2/3$. Also for these densities, we have found that the spin gap region starts at smaller couplings $J/t$ with increasing $\alpha$. The value of the spin gap for $\alpha = 1$ can be seen in Fig. 2 for different densities $n$ as a function of the coupling $J/t$.

## V. CRITICAL EXPONENTS

In 1D, we can describe the long range fluctuations of systems belonging to either of universality classes Tomonaga-Luttinger liquids (TLL) or Luther-Emery liquids (LEL)[22] by a single correlation exponent $K_\rho$.[18,19,22–25] Tomonaga-Luttinger liquids are characterized by gapless charge and spin excitations. The long range correlation functions of TLL decay as power laws. The 1D Hubbard model falls into this universality class,[18,19] and $K_\rho$ decreases for all densities $n \neq 1$ from $K_\rho = 1$ in the free electron limit $U/t \to 0$ to $K_\rho = 0.5$ in the strong coupling limit $U/t \to \infty$. Another example is the $t$-$J$ model before phase separation.[7,8,25] Models with a finite spin gap and have gapless charge excitations generally and belong to the LEL universality class. Charge density wave (CDW) and singlet superconducting correlations have power law behavior, spin density wave (SDW) and triplet superconducting correlations decay exponentially because of the spin gap. Superconducting correlations dominate the long range fluctuations for both universality classes if $K_\rho > 1$. This occurs for the $t$-$J$ model in a narrow region of the phase diagram before phase separation.[7,8]

It is known from conformal field theory that $K_\rho$ can be calculated from numerical results of the low energy spectrum in finite systems alone. The results thus obtained have to be checked for consistency a posteriori. We calculate $K_\rho$ from a combination of two relations. The first one expresses $K_\rho$ by the compressibility $\kappa$ and the charge velocity $v_c$:[23]

$$K_\rho = \frac{\pi v_c n^2 \kappa}{2}. \qquad (9)$$

Since the charge velocity $v_c$ is difficult to determine from the spectrum, we have used a second relation to eliminate $v_c$. The Drude weight $\sigma_0$ of the ac conductivity is given by the energy shift of the ground state in the presence of a field[26] and for 1D systems also by the correlation exponent[18]

$$\sigma_0 = \pi L \frac{\partial^2 E_0(\phi)}{\partial \phi^2}\bigg|_{\phi=0} = 2 v_c K_\rho, \qquad (10)$$

where $E_0(\phi)$ denotes the ground state energy of the system with twisted boundary conditions with a phase factor $\phi$. Together with Eq. (9) we get

$$K_\rho = \sqrt{\frac{\pi}{4} \sigma_0 n^2 \kappa}. \qquad (11)$$

The phase diagrams for $\alpha = 0.4$ and 1 are shown Fig. 3, and we discuss them by comparing with the phase diagram for the $t$-$J$ in Ref. 7. In the limit $J/t \to 0$ the correlation exponent is $K_\rho = 1/2$ for any $\alpha$ and $n \neq 1$ as in the same limit of the $t$-$J$ model. Before phase separation, the value of $K_\rho$ increases monotonically with the coupling $J/t$, and we find as a precursor to phase separation dominant superconducting correlations $K_\rho > 1$. The reason for the strong increase of $K_\rho$ before phase separation is the divergence in the compressibility $\kappa$, as can be seen in Fig. 4 where we show along with $K_\rho$ the compressibility $\kappa$ and the Drude weight $\sigma_0$. Since there is no phase separation for $\alpha = 1$, the superconducting region extends up to $J/t \to \infty$ and the divergence in the compressibility $\kappa$ and in $K_\rho$ is also removed.

The effect of the pair hopping term is different depending on the electron density $n$ and the coupling $J/t$. In the low electron density region we have a strong enhancement of local singlet electron pairs, which can move by the pair hopping term without the cost of breaking a bond $J$. These mobile local electron pairs enhance superconducting correlations and $K_\rho$ is increased with $\alpha$ for all densities $n$ at low couplings $J/t \lesssim 1$. On the other hand, we have a strong suppression of $K_\rho$ with increasing $\alpha$ for low hole densities and couplings $J/t > 1$ and a slight suppression of $K_\rho$ for low electron densities and large $J/t$. The reason for this suppression of $K_\rho$ is that the enhanced propagation of single holes in the low hole density region discussed in Sec. III. This enhancement produces an effective repulsion among holes, which reduces drastically the compressibility $\kappa$, as seen in Fig. 4(b). Therefore the value of the correlation exponent $K_\rho$ via the relation Eq.(11) is also reduced, even though the Drude weight $\sigma_0$ increases with larger values of $\alpha$.

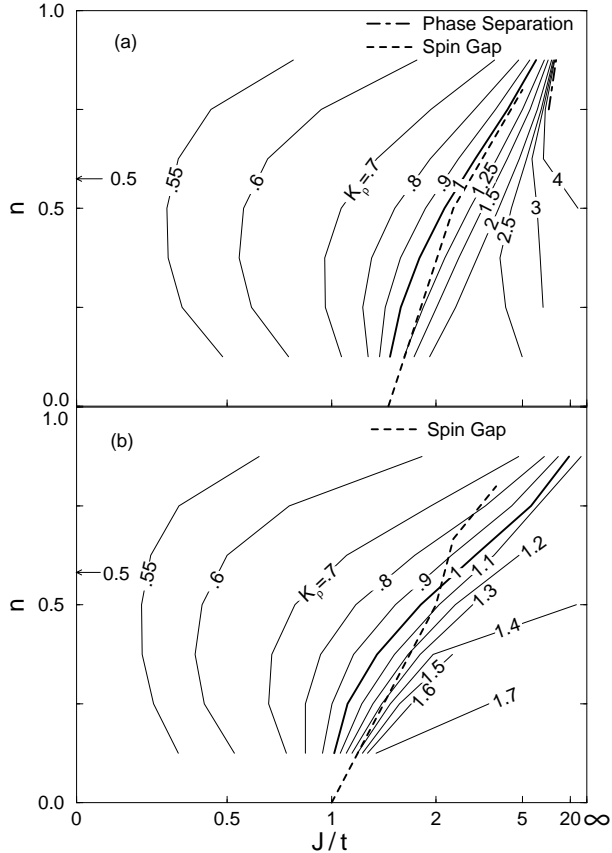

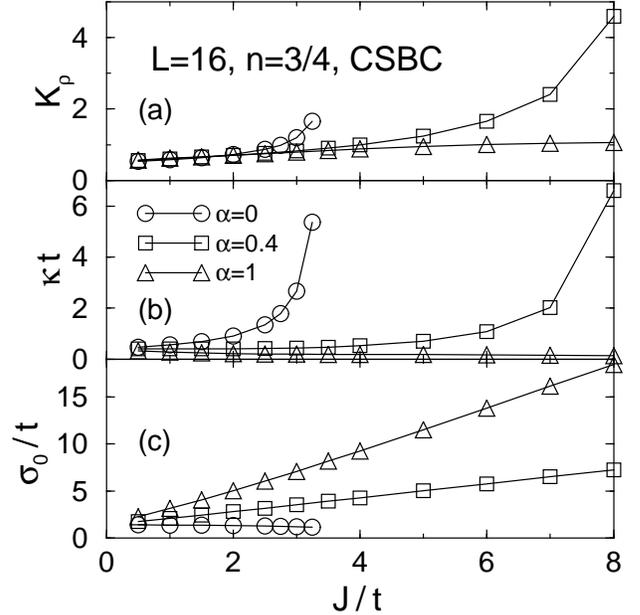

FIG. 3. Phase diagram of the $t$-$J$-$t_3$ model from $L = 16$ cluster calculation. (a) $\alpha = 0.4$, and (b) $\alpha = 1$. The solid curves show the contours of constant correlation exponent $K_\rho$, calculated from for systems with an even number of electrons and CSBC. The dashed-dotted line represents the border line of phase separation, the dashed line the border line of the region of finite spin gap.

FIG. 4. (a) Correlation exponent $K_\rho$, (b) compressibility $\kappa$, and (c) Drude weight $\sigma_0$ for $\alpha = 0, 0.4$ and 1 as a function of $J/t$ at $n = 3/4$. Calculated by exact diagonalization of a lattice of length $L = 16$ and CSBC.

In addition, we have also shown the border line of the spin gap region in the phase diagrams Fig. 3. The contour lines of the correlation exponent seem to be continuous at this transitions line, although the long range behavior of the correlation functions changes drastically.

## VI. CHARGE AND SPIN STRUCTURES

We examine the charge and spin structure factors in more details and calculate them by a QMC method. The charge and spin structure factors are defined as:

$$S_{\text{charge}}(k) = \frac{1}{L} \sum_{j,m}^{L} e^{ik(j-m)} \langle (n_{j,\uparrow} + n_{j,\downarrow})(n_{m,\uparrow} + n_{m,\downarrow}) \rangle,$$

$$S_{\text{spin}}(k) = \frac{1}{L} \sum_{j,m}^{L} e^{ik(j-m)} \langle (n_{j,\uparrow} - n_{j,\downarrow})(n_{m,\uparrow} - n_{m,\downarrow}) \rangle,$$

$$= \frac{4}{L} \sum_{j,m}^{L} e^{ik(j-m)} \langle S_j^z S_m^z \rangle. \qquad (12)$$

According to conformal field theory we may expect $2k_F$ and $4k_F$ CDW fluctuations ($2k_F = n\pi$) in the charge sector and $2k_F$ SDW fluctuations in the spin sector to dominate. The onsite repulsion enhances $4k_F$ CDW fluctuations, whereas the Heisenberg interaction term favors antiferromagnetic alignment of spins on adjacent sites and thus a $2k_F$ SDW state. On the other hand, the formation of bound singlet pairs enhances $2k_F$ CDW fluctuations[5] and favors a maximum at $k = \pi$ in the spin sector.

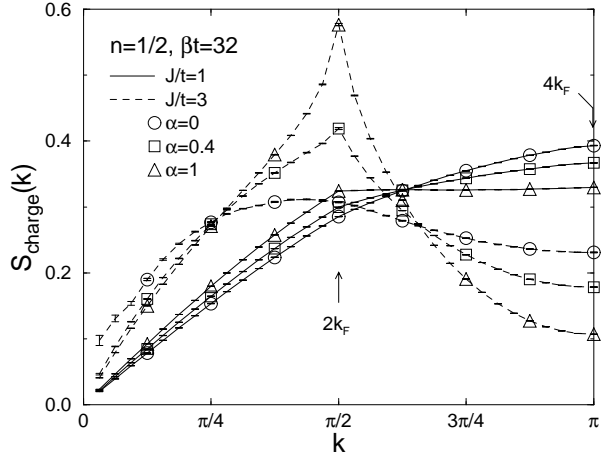

FIG. 5. Charge structure factors of the $t$-$J$-$t_3$ model at $\alpha = 0, 0.4$ and $1$ and $n = 1/2$, calculated by QMC simulation on a lattice of $L = 64, \beta t = 32, \Delta\tau t = 0.25$.

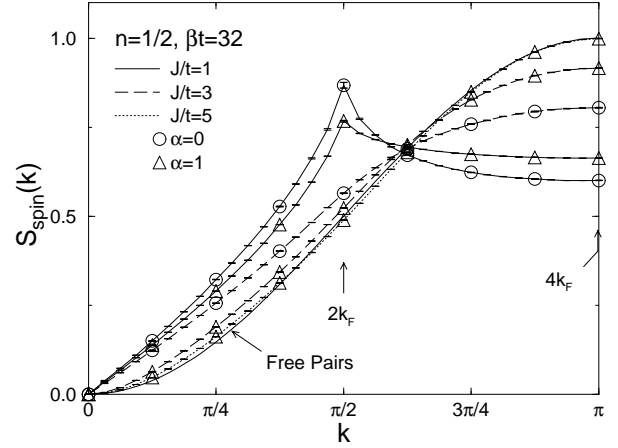

FIG. 6. Spin structure factors of the $t$-$J$-$t_3$ model at $\alpha = 0$ and $1$ and $n = 1/2$, calculated by QMC simulation on a lattice of $L = 64, \beta t = 32, \Delta\tau t = 0.25$. For comparison, the result of a gas of free pairs $S_{\text{spin}}(k) = n(1 - \cos(k))$ is included.

In order to investigate the effect of switching on $\alpha$ for different values $J/t$, we first calculate the charge structure factor. Figure 5 shows the results at $n = 1/2$ for $\alpha = 0, 0.4$ and $1$ for couplings $J/t = 1$ and $3$. Starting at small $J/t$, we expect a cusp at $k = 4k_F$ for $J/t = 1$ and possibly one at $k = 2k_F$ also. There is a maximum at $4k_F (=\pi)$ for all values of $\alpha$, but the cusp at $k = 2k_F$ is hardly visible at $J/t = 1$, since the prefactor of $2k_F$ fluctuations is much smaller than for $4k_F$ fluctuations.[8] Increasing $\alpha$ enhances pairing of electrons and produces stronger $2k_F$ fluctuations. Therefore the cusps at $k = 2k_F$ are much clearer for larger $\alpha$. Pairing is also enhanced by increasing the coupling $J/t$. At the same time, a spin gap opens for the larger values of $\alpha$. This crossover to a LEL is accompanied by a jump in the exponent of the $2k_F$ CDW correlations from $1 + K_\rho$ to $K_\rho$. Accordingly there is a peak at $k = 2k_F$ diverging with the lattice size $L$ in Fig. 5 for $\alpha = 1$ and $J/t = 3$, while the $4k_F$ fluctuations are strongly suppressed.

To analyze the effect of the pair hopping term in the spin sector, we show in Fig. 6 the spin structure factor at $n = 1/2$ for $\alpha = 0$ and $1$ for couplings $J/t = 1, 3$ and $5$. There is a complete spin degeneracy at $J = 0$, which is lifted by an infinitesimal value of $J/t$. The expected $2k_F$ SDW fluctuations can be seen by a clear peak at $k = 2k_F$ at $J/t = 1$ for both values of $\alpha$. With the enhanced formation of pairs by increasing $J/t$, this peak is completely smeared out into a broad maximum at $k = \pi$, reflecting short range nearest neighbor antiferromagnetic fluctuations. For comparison, we have included the structure factor of a gas of free nearest neighbor pairs $S_{\text{spin}}(k) = n(1 - \cos(k))$. There is a good agreement for $\alpha = 1$ and $J/t = 5$ with the calculated values.

In the $t$-$J$ model (that is $\alpha = 0$), the dominant antiferromagnetic configuration is a squeezed spin chain.

The pair hopping term on the other hand favors a local singlet character of the spins. To investigate which effect is dominating for $\alpha \neq 0$, we calculate the correlation of the two spins on both sides of a single hole $C_{\text{shs}} = \langle S_i^z h_{i+1} S_{i+2}^z \rangle / \langle n_i h_{i+1} n_{i+2} \rangle$, and that of the two spins preceding a hole $C_{\text{ssh}} = \langle S_i^z S_{i+1}^z h_{i+2} \rangle / \langle n_i n_{i+1} h_{i+2} \rangle$. The results are shown in Fig. 7 as a function of $J/t$ for $\alpha = 0$ and $1$ with $n = 1/2$ and $3/4$. Since the charge and spin degrees of freedom are decoupled in TLL, we may expect approximately the same values for $C_{\text{shs}}$, $C_{\text{ssh}}$ and $\langle S_i^z S_{i+1}^z \rangle$. We can see in Fig. 7 that this is roughly fulfilled for $\alpha = 0$, as $C_{\text{shs}}$ and $C_{\text{ssh}}$ are of the same order of magnitude. The phase separation boundary, $J/t \simeq 3.3$, for $\alpha = 0$ and $n = 1/2$ can be seen as a cusp in $C_{\mathrm{shs}}$. On the other hand, we may expect $C_{\mathrm{shs}} = 0$ and $C_{\mathrm{ssh}} = -1/4$ for a gas of free singlet bound pairs. Figure 7 shows that $C_{\mathrm{shs}} \to 0$ and $C_{\mathrm{ssh}} \to -1/4$ with increasing $J/t$. This is due to the enhanced pairing for $\alpha = 1$. We can see that $C_{\mathrm{ssh}} \to -1/4$ much faster than $C_{\mathrm{shs}} \to 0$. This means that next to a hole there is almost always a singlet pair allowing a larger kinetic energy by pair hopping, but due to the other terms there is still some antiferromagnetic correlation between two spins separated by a hole.

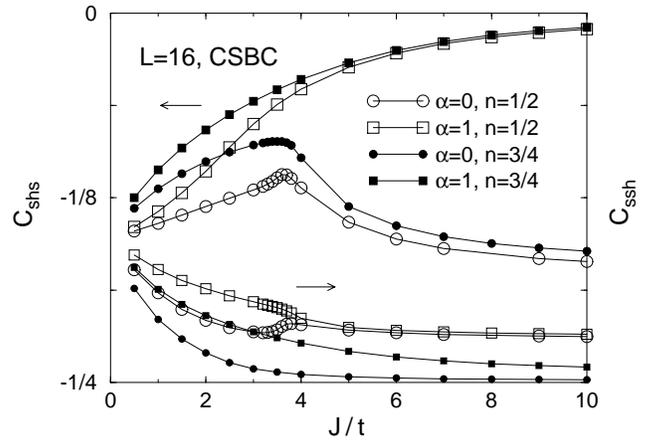

FIG. 7. Three-site correlation of a hole and two spins $C_{\text{shs}} = \langle S_i^z h_{i+1} S_{i+2}^z \rangle / \langle n_i h_{i+1} n_{i+2} \rangle$ and $C_{\text{ssh}} = \langle S_i^z S_{i+1}^z h_{i+2} \rangle / \langle n_i n_{i+1} h_{i+2} \rangle$ in the $t$-$J$ and $t$-$J$-$t_3$ model for different values of $\alpha$, calculated by exact diagonalization of a lattice of length $L = 16$ and CSBC.

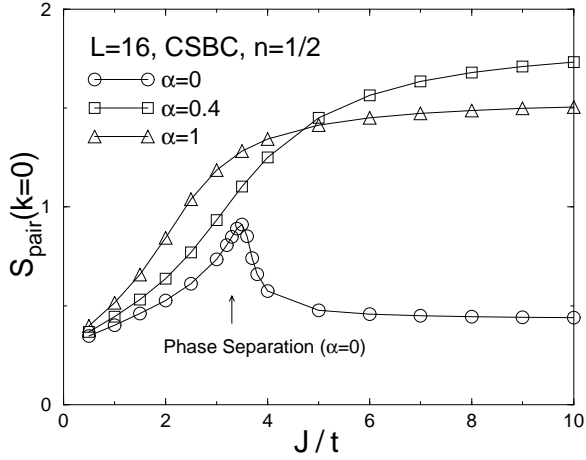

FIG. 8. Pair structure factor $S_{\mathrm{pair}}(k=0)$ for the $t$-$J$-$t_3$ model at $n = 0.5$, calculated by exact diagonalization of a lattice of length $L = 16$ and CSBC.

## VII. SUPERCONDUCTIVITY

Finally we concentrate on details of the superconducting correlations and investigate the suppression of superconductivity in the low hole density region. The singlet superconducting structure factor is defined as

$$S_{\mathrm{pair}}(k) = \frac{1}{L} \sum_{j,m}^{L} e^{ik(j-m)} \langle P_j^\dagger P_m \rangle, \quad (13)$$

where again $P_j^\dagger = (c_{j\uparrow}^\dagger c_{j+1\downarrow}^\dagger - c_{j\downarrow}^\dagger c_{j+1\uparrow}^\dagger)/\sqrt{2}$ is a nearest neighbor singlet pair.

As pointed out in Sec. V, dominating superconducting correlations are expected for $K_\rho > 1$. We can look for signs of enhancement of superconductivity by calculating the uniform component, $S_{\mathrm{pair}}(0)$, which contains both short and long range correlations. In Fig. 8 we have plotted $S_{\mathrm{pair}}(0)$ as a function of $J/t$ for $n = 1/2$ and $\alpha = 0$, 0.4 and 1. For small to intermediate couplings $J/t < 4$ the pair hopping term enhances superconductivity, since $S_{\mathrm{pair}}(0)$ has larger values with increasing $\alpha$.

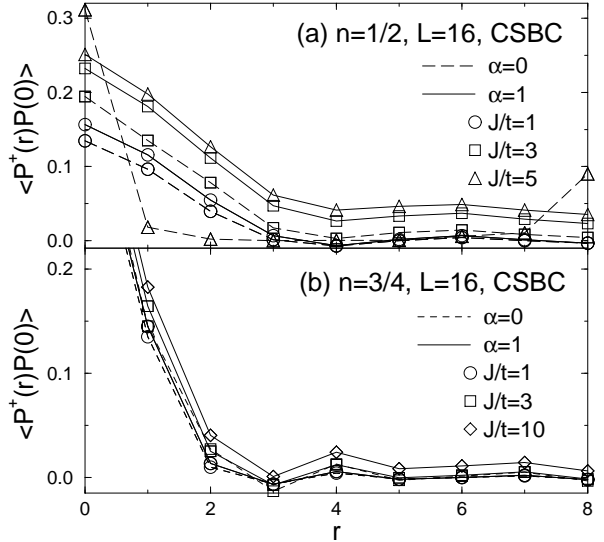

FIG. 9. Real space pairing correlations for the $t$-$J$-$t_3$ model at $\alpha = 0$ and 1. The density is (a) $n = 1/2$ and (b) $n = 3/4$. Calculated by exact diagonalization of a lattice of length $L = 16$ and CSBC.

For very strong couplings $J/t > 4$ an intermediate value of $\alpha = 0.4$ is best among the three considered, producing the largest values of $S_{\mathrm{pair}}(0)$. As there is no long range fluctuations of superconductivity in the phase separated region, we can see a sharp drop in $S_{\mathrm{pair}}(0)$ at the boundary of phase separation, $J/t \simeq 3.5$, for $\alpha = 0$. We have obtained signs of enhanced superconductivity in the same parameter regions indirectly, through the values obtained for the correlation exponent $K_\rho$ in Sec. V.

To separate the long range part of $S_{\mathrm{pair}}(0)$ we need to examine correlations in real space, $\mathcal{P}(r) = \langle P_r^\dagger P_0 \rangle$. We show $\mathcal{P}(r)$ for $\alpha = 0$ and 1 for $J/t = 1, 3$ and 5 in Fig. 9. For $J/t = 1$, superconductivity is not a dominant correlation and $\mathcal{P}(r)$ vanishes for the larger values of $r$. On the other hand, in the superconducting region ($J/t = 3$ for $\alpha = 0$ and $J/t = 3, 5$ for $\alpha = 1$) the correlations $\mathcal{P}(r)$ are quite large for all $r$. The enhancement of superconductivity by the pair hopping term is demonstrated by the long range correlations of $\mathcal{P}(r)$, which are much larger for $\alpha = 1$ than for $\alpha = 0$. The coupling $J/t = 5$ is in the phase separated region for $\alpha = 0$. Therefore $\mathcal{P}(r)$ is only large for $r = 0$ and 8, where a pair of electrons is annihilated and created at either end of the Heisenberg chain.

The suppression of superconductivity for high electron densities, already discussed in Sec. V, can also be seen in the real space pairing correlations $\mathcal{P}(r)$ Fig. 9 for $n = 3/4$. As for $n = 1/2$, the long range parts of $\mathcal{P}(r)$ are larger in the superconducting parameter region ($J/t = 3$ for $\alpha = 0$ and $J/t = 10$ for $\alpha = 1$) than in the non-superconducting region. But also these long range parts of $\mathcal{P}(r)$ are much smaller compared to $n = 1/2$.

Obviously most of the contributions to $S_{\text{pair}}(0)$ originate from short range correlations and superconductivity is suppressed. The reason for the suppression of the long range fluctuations $\langle P_r^\dagger P_0 \rangle$ is the lack of pairs of holes. The pair hopping term produces an effective repulsion among holes, as it has been pointed out in Secs. III and V. Therefore most of the holes are isolated, and the correlation $p_{\text{sh}} = \langle n_i h_{i+1} n_{i+2} \rangle / (1 - n)$ in Tab. I remains close to one for $\alpha = 1$. A strong local singlet nature of electrons allows the annihilation of a spin singlet pair of two neighboring electrons, but it is difficult to find the pair of holes necessary to create another nearest neighbor singlet electron pair, at a distance $r \neq 0$ or 1. This is the main reason of suppression of superconducting correlations even though the pair hopping term enhances a singlet correlation between neighboring electrons.

## VIII. CONCLUSIONS

In this paper we have numerically investigated various aspects of the pair hopping terms in the 1D $t$-$J$-$t_3$ model and compared to the Hubbard and $t$-$J$ models. The pair hopping terms derived from the Hubbard model by the strong coupling expansion in $U/t$ are usually neglected in the $t$-$J$ model. It is found that upon the introduction of these terms, the qualitative similarity of the Hubbard and $t$-$J$ models is improved in the small $J/t$ limit and the values of the correlation exponent $K_\rho$ of the $t$-$J$-$t_3$ and Hubbard models become similar also quantitatively for $J/t < 1$.

The pair hopping terms might be particularly important to superconducting correlations, since singlet electron pairs can hop without breaking using the three site term, therefore a larger region of dominant superconductivity fluctuations is expected. Indeed, our numerical calculations show a complete suppression of phase separation and an extended superconducting region up to $J/t \to \infty$. But the pair hopping terms also favor the propagation of single holes by the larger gain of kinetic energy compared to forming bound hole pairs, which produces an effective repulsion among holes and there are practically no pairs of holes. Hence once a singlet electron pair is annihilated, is is difficult to find the pair of holes needed to create it again at a some distance $r \neq 0$ or 1 from the annihilation and superconductivity is suppressed in the realistic region for superconductivity in the cuprates, which is low hole density and small couplings $J/t \sim 0.3$-$0.4$. But we find $K_\rho > 1$ already at lower values of $J/t$ compared to the $t$-$J$ model in the low electron density region.

Nevertheless we have to emphasize that due to the complete suppression of phase separation, the region of dominant superconducting fluctuations extends up to $J/t \to \infty$, in contrast to many extended $t$-$J$ models and the simple $t$-$J$ model itself, where this region is restricted to a small precursor region of phase separation.

In the single chain models which we examined here, the region of realistic parameters, $J/t \sim 1/3$, is not favorable for superconductivity and the introduction of the three-site $t_3$-term does not change this conclusion. However in other cases, e.g. a 2D lattice or a two-chain ladder, the situation is quite different and the effect of the $t_3$ term in shifting the onsets of pairing and phase separation boundaries may be much more significant. Actually a recent investigation of the $t$-$J$ model including a next nearest neighbor hopping term $t'$ and the pair hopping terms $t_3$ on a $4 \times 4$ lattice indicated enhancement of superconductivity.[27] We believe that it is important to test these observations on larger 2D lattices, where larger correlation lengths can be investigated, and specifically the realistic low hole doping region is of great interest.

## ACKNOWLEDGMENTS

We wish to thank T. M. Rice and D. Würtz for very instructive and helpful discussions. The calculations have been performed on the Cray Y/MP-464 of ETH Zürich and the Parsytec GCel-1/64 of the IPS ETH Zürich. The work was supported by the Swiss National Science Foundation Grant No. NFP-304030-032833 and by an internal grant of ETH-Zürich.